\documentclass[final,5p,times,twocolumn]{elsarticle}
\usepackage{graphicx,amsmath,amssymb,bm,comment}

\usepackage[usenames]{color}
\usepackage{xcolor}

\newcommand{\bvec}[1]{\mathbf{#1}}%makes boldface vectors
\journal{Physics Letters B}

\begin{document} 
 
\title{Satisfying the compressibility sum rule in neutron matter}

\author{Mateusz Buraczynski}
\author{Samuel Martinello}
\author{Alexandros Gezerlis}
\address{Department of Physics, University of Guelph, Guelph, ON N1G 2W1, Canada}

\begin{abstract} 
The static-response function of strongly interacting neutron matter
contains crucial information on this interacting many-particle system, going 
beyond ground-state properties.
In the present work, we tackle this problem with 
quantum Monte Carlo (QMC) approaches 
at several different densities, using both phenomenological forces and 
(for the first time) chiral effective field theory interactions.
We handle finite-size effects via 
self-consistent energy-density functional (EDF) calculations for 8250 particles
in a periodic volume. 
We combine 
these QMC and EDF computations in an attempt to produce a model-independent
extraction of the static response function. Our results are consistent with 
the compressibility sum rule, which encapsulates the limiting behavior
of the response function starting from the homogeneous equation of state, 
without using the sum rule as an input constraint.
Our predictions on inhomogeneous neutron matter can function as benchmarks for 
other many-body approaches, thereby shedding light on
the physics of neutron-star crusts and neutron-rich nuclei. 
\end{abstract} 

%\pacs{21.65.Cd, 26.60.-c, 21.60.Ka, 21.60.Jz}
 
\maketitle 

Neutron matter is a strongly interacting many-body system, involving a cancellation between the 
kinetic energy of the neutrons and the two- and three-neutron potential energy.
Pure neutron matter is an idealization, albeit one that has connections with the nuclear symmetry
energy, experimentally probed via heavy-ion collisions, and with the beta-stable matter that is found
inside neutron stars~\cite{Gandolfi:2015,Lattimer:2016}. Recently, gravitational radiation coming from the merging of two neutron 
stars has been directly detected, leading to a whole array of possible interplay between 
microscopic nuclear physics and its astrophysical implications~\cite{Abbott:2017,Tews:2018,Tong:2020}. On the other hand, 
the strongly correlated nature of neutron matter also gives rise to intriguing connections
with systems composed of ultracold atoms, probed in table-top experiments here on earth~\cite{Gezerlis:2008, Boulet:2018}. 

Neutron-matter properties can be tackled both phenomenologically and from first principles.
In the context of nuclear many-body physics, \textit{ab initio} refers to approaches which 
start with nucleonic degrees of freedom, nucleons exchanging pions, and then computes
properties such as the equation of state (EOS) of neutron matter for a given Hamiltonian~\cite{Friedman:1981,Akmal:1998,Schwenk:2005,Gezerlis:2008,Epelbaum:2009,Kaiser:2012,Friman:2019}.
The latter involves two- and three-nucleon forces which typically contain low-energy couplings
fit to few-body physics~\cite{Carlson:Morales:2003,Gandolfi:2009,Gezerlis:2010,Gandolfi:2012,Baldo:2012,Hebeler:2010,Tews:2013,Gezerlis:2013,Coraggio:2013,Hagen:2014,Gezerlis:2014,Carbone:2014,Roggero:2014,Wlazlowski:2014,Tews:2016,Soma:2020}. In recent decades, there has been a drive in nuclear theory to carry out
such calculations using techniques like, e.g., quantum Monte Carlo 
or Coupled Cluster, which have no free
parameters and therefore provide controlled first-principles approximations. Of course, such 
\textit{ab initio} work is computationally costly and therefore
of somewhat limited applicability; thus, a lot of work is still carried
out using more phenomenological energy-density functional theories of nuclei and infinite matter~\cite{Bender:2003,Chabanat:1998,Rios:2014,Lacroix:2016}.
EDF theories involve a number of parameters which are fit to nuclear masses and radii, but also
to pseudodata coming from \textit{ab initio} many-body theories themselves (such as the EOS of neutron
matter~\cite{Chabanat:1998, Fayans:1998,Brown:2000,Chappert:2008,Fattoyev:2010,Fattoyev:2012,Brown:2014,Rrapaj:2016}, the pairing gap~\cite{Chamel:2008}, the energy of a neutron impurity~\cite{Forbes:2014,Roggero:2015}, or the properties of neutron drops~\cite{Pudliner:1996,Pederiva:2004,Gandolfi:2011,Potter:2014,Gandolfi:2016}).

Much effort has been expended on going beyond the ground-state energy of the many-neutron system.
The theoretical approach is sometimes novel and other times a straightforward extension of the 
ground-state formalism.
As one example, computing the single-particle excitation spectrum of strongly interacting neutron matter
allows for the extraction of the effective mass, which impacts the maximum mass of a neutron star 
as well as the analysis of giant quadrupole resonances~\cite{Li:2018,Isaule:2016,Grasso:2018,Bonnard:2018,Buraczynski:2019,Buraczynski:2020,Bonnard:2020}. Another physical setting, on which
the present Letter is focused, involves placing strongly interacting neutron matter inside a periodic external field.
This gives rise to the static response of neutron matter, a problem which is of intrinsic interest but also
has direct application to neutron-star crusts: there, the deconfined neutrons interact strongly with each other 
and with a lattice of nuclei. 
Both the physics and the theoretical machinery that are relevant here~\cite{Pines:1966} are
analogous to experimentally falsifiable
studies of liquid $^4$He~\cite{Moroni:1992,Moroni:1995} and of cold-atomic systems placed within optical lattices~\cite{Boulet:2018}.

The static-response problem has been tackled with a variety of mean-field-like approaches~\cite{Pastore:2015,Iwamoto:1982,Olsson:2004,Chamel:2012,Chamel:2013,Kobyakov:2013,Pastore:2014,Chamel:2014,Davesne:2015}, but the 
\textit{ab initio} work on the subject has so far been limited~\cite{Buraczynski:2016,Buraczynski:2017}. In the present Letter we study four different densities, using both older nuclear forces (AV8'+UIX)
as well as local N$^2$LO chiral interactions, which are intended to have a closer
connection with the symmetries of the underlying fundamental theory of quarks and gluons. As will
be discussed in more detail below, carrying out such computationally demanding calculations for 
two classes of nuclear interactions, four densities, seven periodicities of the external potential, and
five strength parameters, is already a very challenging task, going well beyond what's been reported
on in the literature in the past. However, in an attempt to ensure that our QMC calculations
can provide sensible extractions of the static response function for neutron matter, we
also report on original calculations we have carried out using a variety of self-consistent 
Skyrme-Hartree-Fock/energy-density functional approaches to the same problem of periodically modulated neutrons placed in a periodic
box. The upshot of combining our QMC and our EDF computations is that we were able to produce
dependable values for the static-response function of neutron matter 
which satisfy the compressibility sum rule (CSR). This, the main result of our work, is shown 
in Fig.~\ref{fig:QMC_resp}.

Let us now provide the briefest of summaries on linear response theory, in order to set up the notation
and also explain why being able to satisfy the CSR should be considered a success.
The linear density-density response function $\chi$ describes the change, up to first order, in density of a system due to a static external field $v$~\cite{Pines:1966,Senatore:1999}:
\begin{align}
\delta n({\bf{r}})=\int \chi({\bf{r'}}-{\bf{r}})v({\bf{r'}}) d^3r',
\label{eq:resp1}
\end{align}
where $n$ is the number density. For a monochromatic potential 
$2v_q{\rm{cos}}({\bf{q}} \cdot {\bf{r}})$ (where the strength is controlled
by $v_q$ and the periodicity by $\mathbf{q}$) the change in energy per particle is given by
\begin{align}
\Delta E/N=\frac{\chi(q)}{n_0}v_q^2+\sum_{i=2}^{\infty}C_{2i}v_q^{2i},
\label{eq:respfit}
\end{align}
where $n_0$ is the unmodulated density, the $C_{2i}$ are related to higher-order response functions, 
and we are now dealing with the response
function in wave-number space, $\chi(q)$. 
By combining the Kramers-Kronig relations with the fluctuation-dissipation theorem, one arrives
at a way of relating the static-response function with an integral over energy involving the dynamical structure factor $S(q,\omega)$~\cite{Pines:1966,Senatore:1999}. At zero temperature this is:
\begin{align}
\chi(q)=-\frac{n_0}{\pi\hbar}\int_0^{\infty}d\omega \frac{S(q,\omega)}{\omega}
\end{align}
In the limit $q\to 0$ this yields the compressibility sum rule:
\begin{align}
\frac{1}{\chi(0)}&=\frac{1}{n_0}\Bigg (\frac{\partial p}{\partial n_0}\Bigg	 )_{T=0}=\frac{\partial^2(n_0 E/N)}{\partial n_0^2}\Bigg|_{T=0}
\label{eq:comp}
\end{align}
where $p$ is the pressure.
This connects the linear response at $q=0$ to the equation of state of the homogeneous
gas. In other words, the EOS provides a constraint on the response at $q=0$.

While the theory of linear response, and the CSR, are completely general results, we now turn
to our specific problem, characterized by the Hamiltonian:
\begin{align}
\hat{H}=\Bigg\{-\frac{\hbar^2}{2m}\sum_i \nabla_i^2+V_{\rm{ext}}\Bigg\}+\Bigg\{\sum_{i<j}V_{ij}+\sum_{i<j<k}V_{ijk}\Bigg\},
\label{eq:H}
\end{align}
which we have separated into one-body and more-body terms. The
static response induced by ``turning on" the one-body term $V_{\rm{ext}}=\sum_i2v_q{\rm{cos}}({\bf{q}} \cdot {\bf{r}}_i)$ (via increasing $v_q$ from zero) allows one to extract the Fourier component of the linear density-density response function $\chi(q)$ via Eq.~(\ref{eq:respfit}).  With a view
to capturing the sensitivity of the static response on detailed features of the microscopic two-
and three-nucleon forces, here we employ:
(i) the high-quality phenomenological potential Argonne v8'~\cite{Wiringa:2002} 
together with the Urbana IX potential~\cite{Pudliner:1997}, and (ii)
local chiral forces at next-to-next-to-leading order (N$^2$LO), namely the $R_0~=~1.0$ fm two-nucleon interaction of Ref.~\cite{Gezerlis:2014} and the 
$R_{3N}~=~1.0$ fm three-nucleon interaction of Ref.~\cite{Tews:2016}.

For a given Hamiltonian, we carry out quantum Monte Carlo computations~\cite{Lynn:2019}. Specifically,
we begin with a variational Monte Carlo calculation, which evaluates the expectation value
of the Hamiltonian for a given trial state vector $|\Psi_T\rangle$. This is followed
by an Auxiliary Field Diffusion Monte Carlo (AFDMC) calculation; 
AFDMC is a projector method that extracts the ground state from $|\Psi_T\rangle$, 
by propagating forward in imaginary time~\cite{Pudliner:1997}, via 
a stochastic evolution over a set of configurations of the particle positions and spins~\cite{Schmidt:1999}. 
The ground-state energy is an average over the evolved configurations:
\begin{align}
E_0=\frac{1}{M}\sum_{i=1}^M\frac{\langle{\bf R}_i,{\bf S}_i|\hat{H}|\Psi_T\rangle}{\langle{\bf R}_i,{\bf S}_i|\Psi_T\rangle}
\label{eq:AFDMC_Energy}
\end{align}
The calculations scale as the number of particles $N$ cubed, limiting one to about $100$ particles. 
As is commonly done, we employ $66$ particles in a periodic box.
Even at this early stage, it is worth emphasizing that the choice of 66 particles comes from outside 
the \textit{ab initio} many-body framework~\cite{Hagen:2014,Gezerlis:2014}, 
i.e., from the simplest possible theory, namely that of a non-interacting Fermi gas.
(We return to this below.)

\begin{figure}[t]
\begin{center}
\includegraphics[width=1.0\columnwidth]{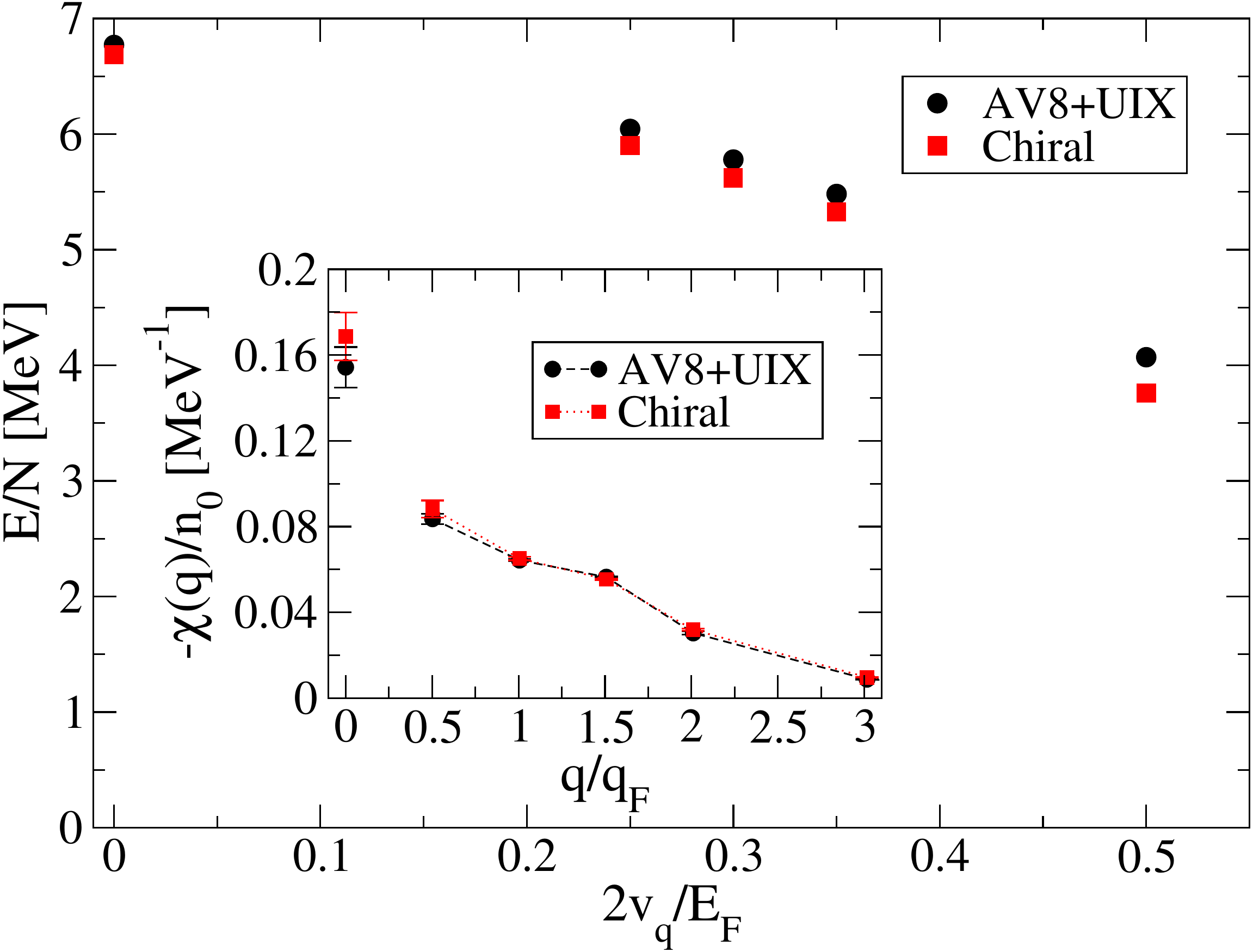}
\caption{AFDMC energy per particle of $66$ particles versus strength of the one-body potential,
for two choices of nuclear interactions. The particle density is $0.04\, \rm{fm}^{-3}$ and one period fits inside the box. Inset: the corresponding response functions, extracted 
assuming a Mathieu-based finite-size prescription.
The isolated points at $q=0$ are the CSR constraints.
\label{fig:Evq}}
\end{center}
\end{figure}

To begin with, we focus on a density of $0.04\, \rm{fm}^{-3}$: note that $q$ controls
how many periods of the external potential fit inside the box of 66 neutrons, whereas $v_q$ 
is the strength of that potential. In Fig.~\ref{fig:Evq} we show the case where a single period
of the external cosine potential fits in the box for five $v_q$ values for each choice of
the nuclear force; $v_q=0$ corresponds to the unmodulated (homogeneous) case.
Qualitatively, we find a greater curvature for the chiral interaction compared with AV8'+UIX, implying
a larger response. This is borne out by the inset of Fig.~\ref{fig:Evq} which,
still at a density of $0.04\, \rm{fm}^{-3}$, shows an extraction of 
the response function $\chi(q)$ via Eq.~(\ref{eq:respfit}) truncated to the first two terms.

The inset also contains isolated points at $q=0$, which were obtained using the 
compressibility sum rule
from Eq.~(\ref{eq:comp}). Not only are the response functions somewhat ragged, but 
the trend exhibited by the finite-$q$ responses
appears to be inconsistent with the CSR: it is hard to believe that the value of
the response function doubles from $q/q_F \approx 0.5$ to $q=0$. 
In order to make progress, we now note that the results in the inset to Fig.~\ref{fig:Evq} 
are not given for 66 particles. As is customary in QMC studies,
a finite-size prescription has been
employed to go to the thermodynamic limit (TL), i.e., to go from 66 to infinitely many particles at constant density:
\begin{align}
\Delta\bar{E}(TL)=\Delta\bar{E}(66) - \Delta\bar{E}_{NI}(66)+\Delta\bar{E}_{NI}(TL)
\label{eq:pres0}
\end{align}
where the bar notation means energy per particle. $E_{NI}$ is the energy of non-interacting fermions with the same external potential. Basically, this amounts to using only the terms in the first
bracket in the Hamiltonian of Eq.~(\ref{eq:H}), i.e., the kinetic energy together with the external cosine potential. 
This is a standard problem giving rise to Mathieu functions and characteristic values. 

To reiterate, all QMC studies of infinite matter proceed with a finite number of particles placed in a periodic
volume. In most studies over the last decade, 66 particles are used, because that is a shell closure
for the Fermi-gas problem of particles that are both free and non-interacting (i.e., have no interactions
with external fields or with each other, respectively). The next step from there is to introduce (either a more elaborate treatment of the boundary
conditions, or) 
a finite-size prescription that is aware of the external field while still neglecting interactions between the
particles; this is the Mathieu-based prescription of Eq.~(\ref{eq:pres0}). Motivated by the mismatch
between the CSR and the QMC extraction of $\chi(q)$ shown in Fig.~\ref{fig:Evq},
we decided to take a further step, trying to account for the impact of the interparticle interactions on the
finite-size effects. To do this, we need a nuclear theory that can handle both finite
systems and infinite matter, but that is also easier to handle computationally. In other words,
the idea is to use a non-QMC theory as guidance on the finite-size effects \textit{only};
as was stressed earlier, this is a step that is already carried out in other QMC studies, though typically
without being spelled out.

\begin{figure}[t]
\begin{center}
\includegraphics[width=1.0\columnwidth]{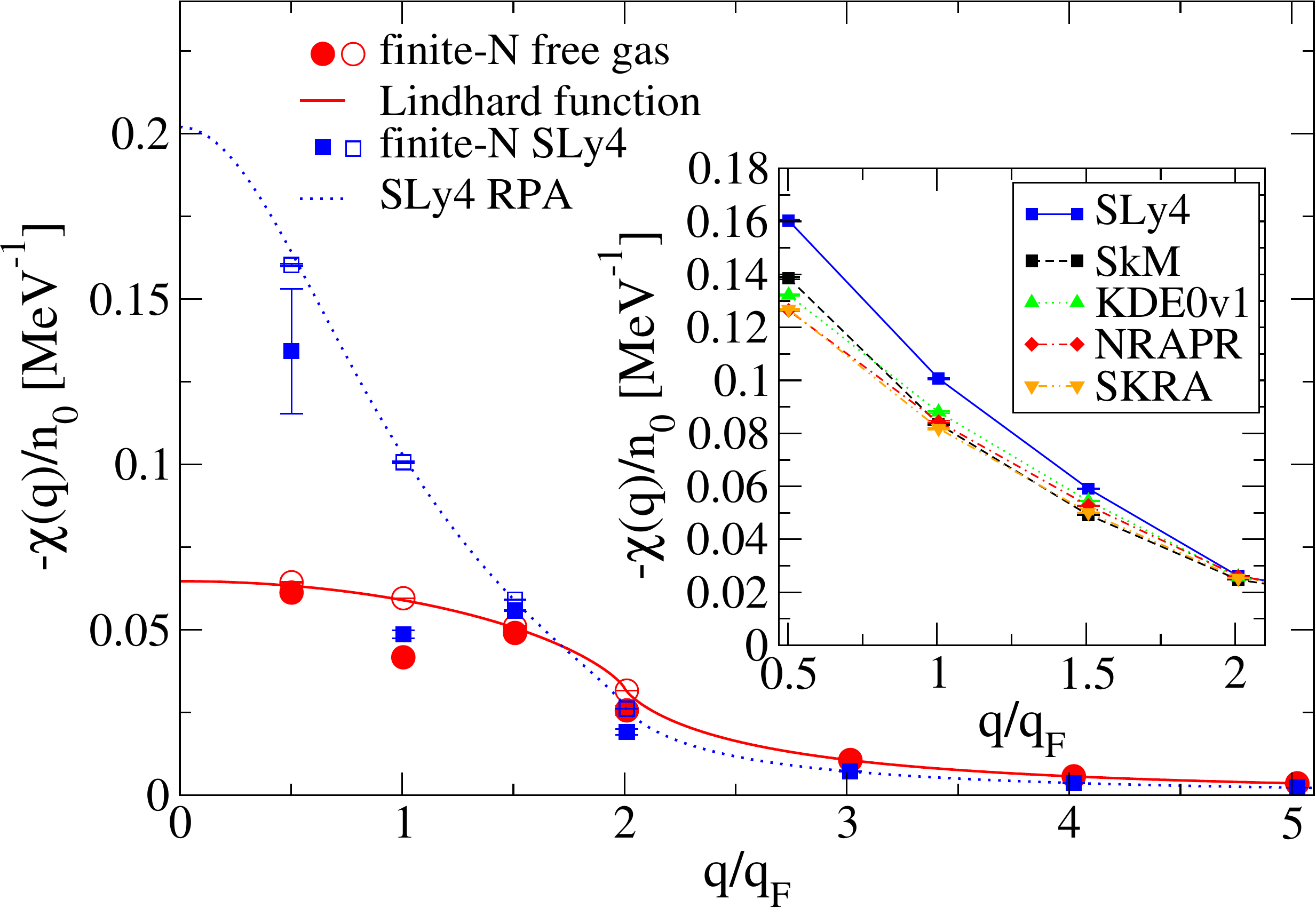}
\caption{Linear static-response functions at a density of $0.04\, \rm{fm}^{-3}$. The circles and squares correspond to non-interacting gas and SLy4 responses, respectively. The solid and dotted lines are TL responses of the non-interacting gas~\cite{Kittel:1969} and SLy4, respectively. Solid symbols correspond to $66$ particles and hollow symbols are for $8250$ particles. Inset: the $8250$-particle response for several Skyrme parametrizations.
\label{fig:resp_ex}}
\end{center}
\end{figure}

Given the widespread adoption and applicability of energy-density functionals, we opted to use
a Skyrme EDF to carry out our finite-size-oriented studies. Specifically,
without loss of generality, we are free to take a coordinate system such that
$\mathbf{q}$ points along the $z$ direction, leading to an external potential of
the form $v(z)$. In that case, the Hartree-Fock
formalism will lead to the following equation:
\begin{align}
	&\frac{\frac{d^2}{d x^2}\phi_{i,x}(x)}{\phi_{i,x}(x)} 
	+ \frac{\frac{d^2}{d y^2}\phi_{i,y}(y)}{\phi_{i,y}(y)} 
	+ \frac{\frac{d^2}{d z^2}\phi_{i,z}(z)}{\phi_{i,z}(z)} \nonumber\\
	&+ \Bigg(\frac{\frac{d}{d z}\frac{\hbar^2}{2m^*(z)}}{\frac{\hbar^2}{2m^*(z)}} \Bigg)\frac{\frac{d}{d z}\phi_{i,z}(z)}{\phi_{i,z}(z)} 
	-\frac{U(z) + v(z) - e_i}{\frac{\hbar^2}{2m^*(z)}} = 0 
	\label{eq:separatedpde}
\end{align}
where
$\phi_i(\mathbf{r}) = \phi_{i,x}(x)\phi_{i,y}(y)\phi_{i,z}(z)$ is
a single-particle orbital and $U(z)$ contains various
Skyrme parameters and densities (something similar holds for 
the effective mass $m^*(z)$).
The implementation details are discussed in 
a forthcoming publication~\cite{Martinello:2020}.
What should be clear here is that we are tackling the Skyrme-Hartree-Fock
static-response problem by combining two non-interacting problems (in the $x$ and $y$ directions)
with an interacting problem in the $z$ direction for the eigenvalues $e_i$
and the eigenfunctions $\phi_{i,z}(z)$. These three problems are solved self-consistently.
As a check, we can switch off the interactions
and recover the Mathieu-based problem.

For the sake of concreteness, we start by determining the energies (and, from Eq.~(\ref{eq:respfit}),
also the responses) for SLy4, a standard Skyrme parametrization. Since our new code works with
periodic boundary conditions and $q$ and $v_q$ at our disposal, we can pick the particle number to 
be either 66 (which would correspond to the QMC studies) or much larger than that. 
For the latter case, we pick a particle number of $66 \times 5^3 = 8250$: this corresponds to a box length that is 
five times larger than that of 66 particles, making it possible to avoid ``stretching'' the external potential.
It goes without saying that it will be impossible to handle 8250 fermions in the context of QMC in the foreseeable future, which is why we employ EDFs for these investigations. 
In Fig.~\ref{fig:resp_ex} we show the TL response for SLy4 as a (dotted) curve and the results
for 8250 particles with hollow squares: we find a very good match. 
The TL response for SLy4 is denoted by RPA, i.e., random-phase approximation (see, e.g., Ref.~\cite{Pastore:2015}).
(We have checked
that results with $66 \times 4^3 = 4224$ particles are essentially identical, except for a very small
difference at the first point; thus, the systematic errors in our large-$N$ computations are well under control). On the other hand, the response for 66 particles is quite different, exhibiting a major dip
at $q \approx q_F$; we would get a third type of behavior if we used another (small) number of particles. 
We also took the opportunity to turn off the interactions, giving rise to the Mathieu-based problem:
there, too, 8250 particles do an excellent job of capturing the TL. Just like in the SLy4 case, 
for 66 particles we find a dip at $q \approx q_F$ 
in comparison to the TL response (known as the Lindhard function~\cite{Kittel:1969}). 
Crucially, this dip is much smaller in size than the corresponding SLy4 one. Another difference in 
SLy4 vs Mathieu behavior has to do with the $q \approx q_F/2$ point: 
for this case, the 66-particle answer for the non-interacting gas is right on top of the TL curve, implying
no finite-size effects (but things are different for SLy4).

To reiterate, we can carry out SLy4 calculations of the response function for 66 and 8250 particles: 
the latter choice matches the thermodynamic limit, whereas the difference between the former and
the latter will help guide the finite-size extrapolation employed in our QMC studies. In equation form, our
finite-size prescription is now:
\begin{align}
\Delta\bar{E}(TL)=\Delta\bar{E}(66) - \Delta\bar{E}_{Sk}(66)+\Delta\bar{E}_{Sk}(8250)
\label{eq:pres1}
\end{align}
instead of Eq.~(\ref{eq:pres0}). 
Of course, there is nothing special about the specific Skyrme
parametrization of SLy4. With a view to making our finite-size prescription as model-independent as possible,
we decide to employ several other Skyrme parametrizations, the response of which is shown in the 
inset to Fig.~\ref{fig:resp_ex}.
SLy4 and SkM* are commonly used,
while NRAPR, SKRA, and KDE0v1 are among a select few parametrizations that respect a set of constraints given by properties of neutron matter~\cite{Brown:2014,Dutra:2012}. 
We capture the 
variation between the parametrizations by averaging out the $\Delta\bar{E}_{Sk}(N)$ terms over all of the functionals, determining the error bar by the spread of the five quantities. 
It's important to note that in Eq.~(\ref{eq:pres1}) 
we are both subtracting and adding in a Skyrme-based quantity: this prescription
addresses only the finite-size effects, i.e., it does \textit{not} bias the response to look like 
that of our Skyrme functional of choice (or place too much trust on the latter).
Thus, while the Mathieu-based finite-size corrections know nothing about 
the interaction, our average in Eq.~(\ref{eq:pres1}) addresses the effect of the interaction on such
corrections.

\begin{figure}[t]
\begin{center}
\includegraphics[width=0.9\columnwidth]{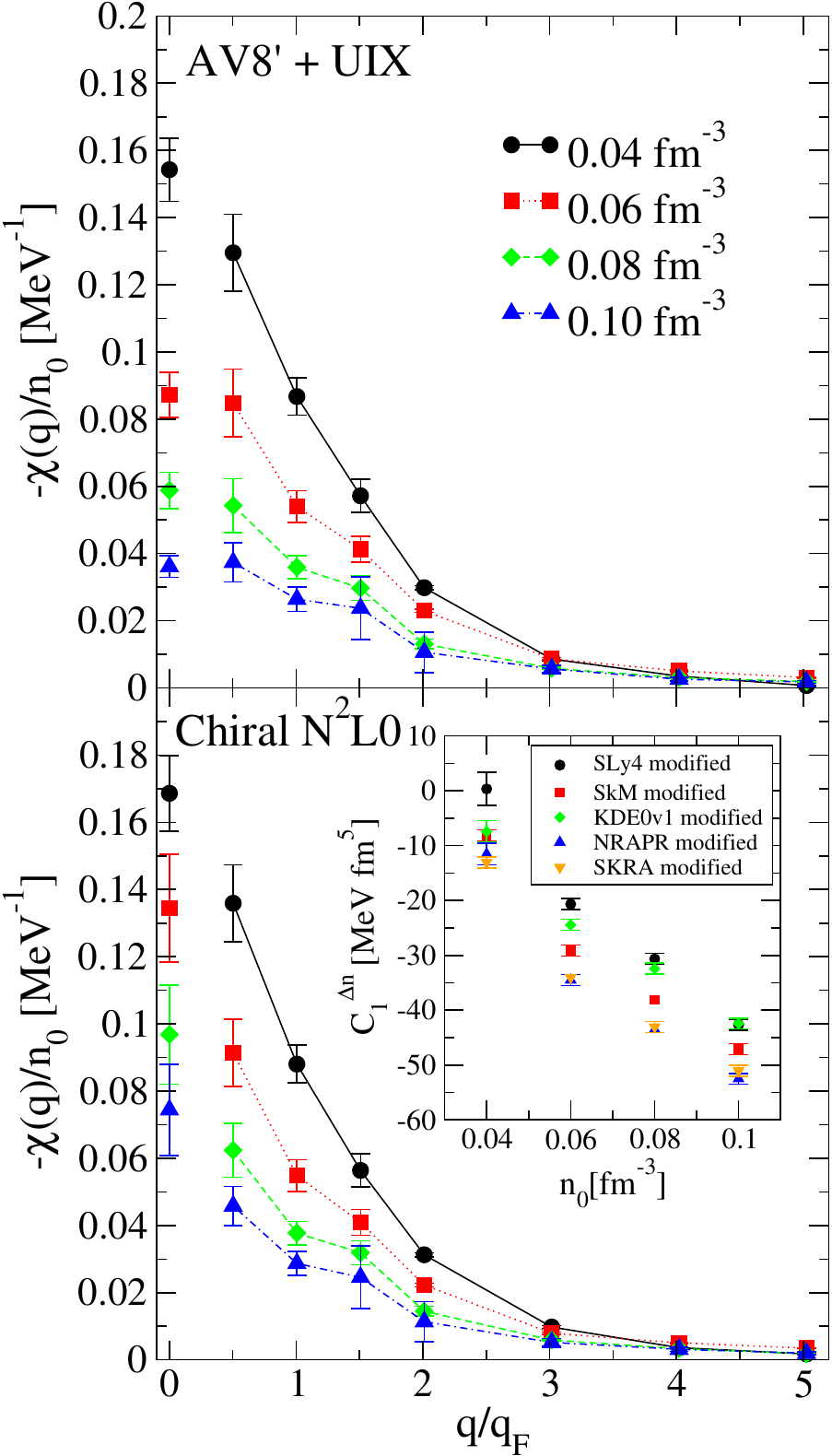}
\caption{AFDMC responses extrapolated via the EDF-based prescription at several densities for AV8'+UIX (top) and 
local N$^2$LO chiral interaction (bottom). The isolated points at $q=0$ are CSR results. 
Inset: modified isovector gradient coefficient for chiral-interaction results.
\label{fig:QMC_resp}}
\end{center}
\end{figure}

Armed with our AFDMC machinery and the EDF-based finite-size prescription of 
Eq.~(\ref{eq:pres1}), we are now ready to turn to our final extractions of the static-response 
function in neutron matter. We study the 
four densities $0.04$, $0.06$, $0.08$, and $0.1\, \rm{fm}^{-3}$ (higher
densities lie in the neutron-star core); 
we also work with 1, 2, 3, 4, 6, 8, and 10 periods inside the box yielding seven response points in total
(as well as five different strengths of the external potential for each case).
The final results are shown in Fig.~\ref{fig:QMC_resp}, for both AV8'+UIX
and N$^2$LO chiral interactions. Overall, we find that the response for the chiral interaction
is larger than that for AV8'+UIX. Our final results also include modest error bars
(reflecting the QMC shape and the spread among EDF predictions), 
which \textit{de facto}
eliminate the raggedness that was seen in Fig.~\ref{fig:Evq}. 
Also, the dip seen in Fig.~1, resulting from the Mathieu-based corrections of Fig.~2, becomes less
pronounced when employing the Skyrme-based prescription (again, compare the distance between solid and hollow points
in Fig.~2), so
that impacts the finite-size corrected QMC results.

Even more importantly, 
we find that our values (connected by lines to guide the eye) do an excellent job of respecting
the compressibility sum rule results, shown as isolated 
points at $q=0$. (Both here and in the inset to Fig.~\ref{fig:Evq}, the CSR
values are insensitive to the finite-size extrapolation scheme employed.)
Crucially, the CSR values were never built-in as input in any form:
we simply took our AFDMC energy results, extrapolated them to the thermodynamic limit using original
EDF calculations, and the final answers end up agreeing with the compressibility sum rule AFDMC 
values: this is a non-trivial
consistency check. (There appears to be a slight density dependence in how well the CSR is satisfied, 
but the error bars make it hard to quantify such a claim.)

Returning to the EDF formalism, we note that 
one way to write down the Skyrme functional is in the isospin representation:
\begin{equation}
{\cal E}_{Sk} = \sum_{T=0,1} \left [ ( C^{n,a}_T + C^{n,b}_T n^{\sigma}_0  ) n^2_T 
 + C^{\Delta n}_T (\nabla n_T)^2 + C^{\tau}_T n_T \tau_T  \right ]
\label{eq:Skyrme}
\end{equation}
where we are only showing the interaction part. For pure neutron systems
$n_0=n_1=n$ and, similarly, $\tau_0=\tau_1=\tau$, where these are
the number density and kinetic density expressed in terms of the single-particle orbitals
introduced after Eq.~(8):
\begin{align}
	n(\bvec{r}) &= \sum_{i}|\phi_i(\bvec{r})|^2 \nonumber \\
	\tau(\bvec{r}) &= \sum_{i}|\bvec{\nabla}\phi_i(\bvec{r})|^2
	\label{eq:tau}
\end{align}
%It is straightforward to translate between these $C_T$ parameters
%and the $t_i$'s and $x_i$'s used above. 
The standard values for the isovector gradient coefficient are
$C^{\Delta n}_1= -15.7, -17.1, -11.4, -16.5, \rm{and}\,-17.1\,\rm{MeV \,fm^5}$ in SLy4, SkM*, KDE0v1, NRAPR, and SKRA respectively.
To produce the inset to Fig.~3, we have adjusted $C^{\Delta n}_1$ for each functional,
leaving all other parameters frozen.
In essence, what this is doing is trying to get 
the EDFs to match the static response of neutron matter given by quantum Monte Carlo.
In other words, it is an attempt to get the EDFs to match not the energies given by QMC,
but the energy differences between the homogeneous and inhomogeneous cases, as defined
in Eq.~(2). 
While these are energetics-focused computations, involving particles in a periodic box,
they are consistent with results following from Skyrme-RPA, as seen in Fig.~2. 
As shown by Eq.~(2), any use of QMC results as synthetic data needs to employ energy
\textit{differences} since these reflect the \textit{response} properties; recall that the QMC and Skyrme 
predictions are already different for the homogeneous/bulk energy case, so this fact needs to be taken
into account
when computing response functions.

Crucially, the results of this plot are the only ones in existence that use chiral EFT interactions as input
to an \textit{ab initio} many-nucleon calculation (which then functions as ``synthetic data'').
The modified gradient coefficient 
impacts the transition density for a neutron star's core-crust boundary~\cite{Lim:2017}.
Crucially, this plot is showing that the isovector gradient coefficient that results from combining
QMC with EDF (keeping the other coefficients frozen, as noted above) is density dependent; this is consistent with what results from the density-matrix expansion~\cite{Buraczynski:2016,Buraczynski:2017,Kaiser:2012b}.

In summary, we have studied periodically modulated neutron matter using two many-body approaches. We 
employed the \textit{ab initio} Auxiliary Field Diffusion Monte Carlo technique 
to obtain energies for 66 particles, using two families of nuclear interactions. 
By also introducing a Skyrme energy-density-functional-based finite-size extrapolation scheme,
we extracted the static-response function for neutron matter at a number of densities. 
Our final results satisfy the compressibility sum rule, implying consistency between our
investigations of inhomogeneous neutron 
matter and our independently carried out studies of homogeneous matter. This is a non-trivial
accomplishment, since it is much easier
to simulate the thermodynamic limit for a homogeneous gas than for a collection of (perhaps isolated)
neutron drops.
In the future, when other many-body techniques are used to tackle the static-response problem, these results
can be used as microscopic benchmarks. Furthermore, since the EDF-based extrapolation scheme 
proposed and employed here 
is not QMC-specific, it could also be used to guide investigations that make use of other \textit{ab initio} methods.

\begin{comment}
At zero temperature there is no heat transfer so pressure is given by
\begin{align}
p&=-\Bigg(\frac{\partial E}{\partial V}\Bigg)_{T=0}\nonumber\\
&=-\Bigg(\frac{\partial E}{\partial n}\Bigg)_{T=0}\Bigg(\frac{\partial n}{\partial V}\Bigg)\nonumber\\
&=\frac{N}{V^2}\Bigg(\frac{\partial E}{\partial n}\Bigg)_{T=0}\nonumber\\
&=n^2\Bigg(\frac{\partial (E/N)}{\partial n}\Bigg)_{T=0}
\label{eq:pressure}
\end{align}
where the second line is an application of the chain rule and the third line is from $n=N/V$.
Rearranging Eq.~\ref{eq:comp} at zero temperature and using Eq.~\ref{eq:pressure}:
\begin{align}
\frac{1}{\chi_n(0)}&=\frac{1}{n}\Bigg (\frac{\partial p}{\partial n}\Bigg	 )_{T=0}\nonumber\\
&=\frac{1}{n} \Bigg[2n\Bigg(\frac{\partial (E/N)}{\partial n}\Bigg)_{T=0}+n^2\Bigg(\frac{\partial^2 (E/N)}{\partial n^2}\Bigg)_{T=0}\Bigg]\nonumber\\
&=2\Bigg(\frac{\partial (E/N)}{\partial n}\Bigg)_{T=0}+n\Bigg(\frac{\partial^2 (E/N)}{\partial n^2}\Bigg)_{T=0}\nonumber\\
&=\frac{\partial}{\partial n}\Bigg[ (E/N)+n\frac{\partial (E/N)}{ \partial n}\Bigg]_{T=0}\nonumber\\
&=\frac{\partial^2(nE/N)}{\partial n^2}\Bigg|_{T=0}
\label{eq:comp2}
\end{align}
where the fourth and fifth lines are obtained by integrating the expression on the previous line and using integration by parts for the second term.
\end{comment}

This work was supported by the Natural Sciences and Engineering Research Council (NSERC) of Canada, the
Canada Foundation for Innovation (CFI), and the Early
Researcher Award (ERA) program of the Ontario Ministry of Research, Innovation and Science. Computational resources were provided by SHARCNET and NERSC.

\end{document}